\documentclass[sigconf]{acmart}

\usepackage{balance}

\def\BibTeX{{\rm B\kern-.05em{\sc i\kern-.025em b}\kern-.08emT\kern-.1667em\lower.7ex\hbox{E}\kern-.125emX}}

\usepackage{booktabs}
\usepackage{multirow}
\usepackage{subfigure}
\usepackage{verbatim}
\usepackage{algorithm}
\usepackage{algorithmic}
\usepackage{makecell}
\usepackage{CJKutf8}
\usepackage{flushend}
\usepackage[utf8]{inputenc}
\copyrightyear{2019} 
\acmYear{2019} 
\setcopyright{acmcopyright}
\acmConference[KDD '19]{The 25th ACM SIGKDD Conference on Knowledge Discovery and Data Mining}{August 4--8, 2019}{Anchorage, AK, USA}
\acmBooktitle{The 25th ACM SIGKDD Conference on Knowledge Discovery and Data Mining (KDD '19), August 4--8, 2019, Anchorage, AK, USA}
\acmPrice{15.00}
\acmDOI{10.1145/3292500.3330665}
\acmISBN{978-1-4503-6201-6/19/08}

\author{Chuhan Wu}
\affiliation{%
  \institution{Tsinghua University}
    \city{Beijing}
  \state{China}
}
\email{wuch15@mails.tsinghua.edu.cn}

\author{Fangzhao Wu}
\affiliation{%
  \institution{Microsoft Research Asia}
  \city{Beijing}
  \state{China}
  \postcode{100080}
}
\email{wufangzhao@gmail.com}

\author{Mingxiao An}
\affiliation{%
  \institution{USTC}
  \city{Hefei}
  \state{Anhui}
}
\email{anmx@mail.ustc.edu.cn}

\author{Jianqiang Huang}
\affiliation{%
  \institution{Peking University}
  \city{Beijing}
  \state{China}
}
\email{1701210864@pku.edu.cn}

\author{Yongfeng Huang}
\affiliation{%
  \institution{Tsinghua University}
    \city{Beijing}
  \state{China}
}
\email{yfhuang@tsinghua.edu.cn}

\author{Xing Xie}
\affiliation{%
  \institution{Microsoft Research Asia}
  \city{Beijing}
  \state{China}
  \postcode{100080}
}
\email{xingx@microsoft.com}

\begin{document}
\title{NPA: Neural News Recommendation with\\ Personalized Attention}

\begin{abstract}
News recommendation is very important to help users find interested news and alleviate information overload.
Different users usually have different interests and the same user may have various interests.
Thus, different users may click the same news article with attention on different aspects.
In this paper, we propose a neural news recommendation model with personalized attention (NPA). 
The core of our approach is a news representation model and a user representation model.
In the news representation model we use a CNN network to learn hidden representations of news articles based on their titles.
In the user representation model we learn the representations of users based on the representations of their clicked news articles.
Since different words and different news articles may have different informativeness for representing news and users, we propose to apply both word- and news-level attention mechanism to help our model attend to important words and news articles.
In addition, the same news article and the same word may have different informativeness for different users. 
Thus, we propose a personalized attention network which exploits the embedding of user ID to generate the query vector for the word- and news-level attentions.
Extensive experiments are conducted on a real-world news recommendation dataset collected from MSN news, and the results validate the effectiveness of our approach on news recommendation.
\end{abstract}

\keywords{News Recommendation, Neural Network, Personalized Attention}

\maketitle

\section{Introduction}
Online news platforms such as MSN News and Google News have attracted a huge number of users to read digital news~\cite{das2007google,lavie2010user}.
However, massive news articles are emerged everyday, and it is impractical for users to seek for interested news from a huge volume of online news articles~\cite{phelan2011terms,wu2019neural}.
Therefore, it is an important task for online news platforms to target user interests and make personalized news recommendation~\cite{ijntema2010ontology,de2012chatter,an2019neuralnews}, which can help users to find their interested news articles and alleviate information overload~\cite{wang2018dkn,wu2019neuralnews}.

\begin{figure}[!t]
  \centering
    \includegraphics[width=1.0\linewidth]{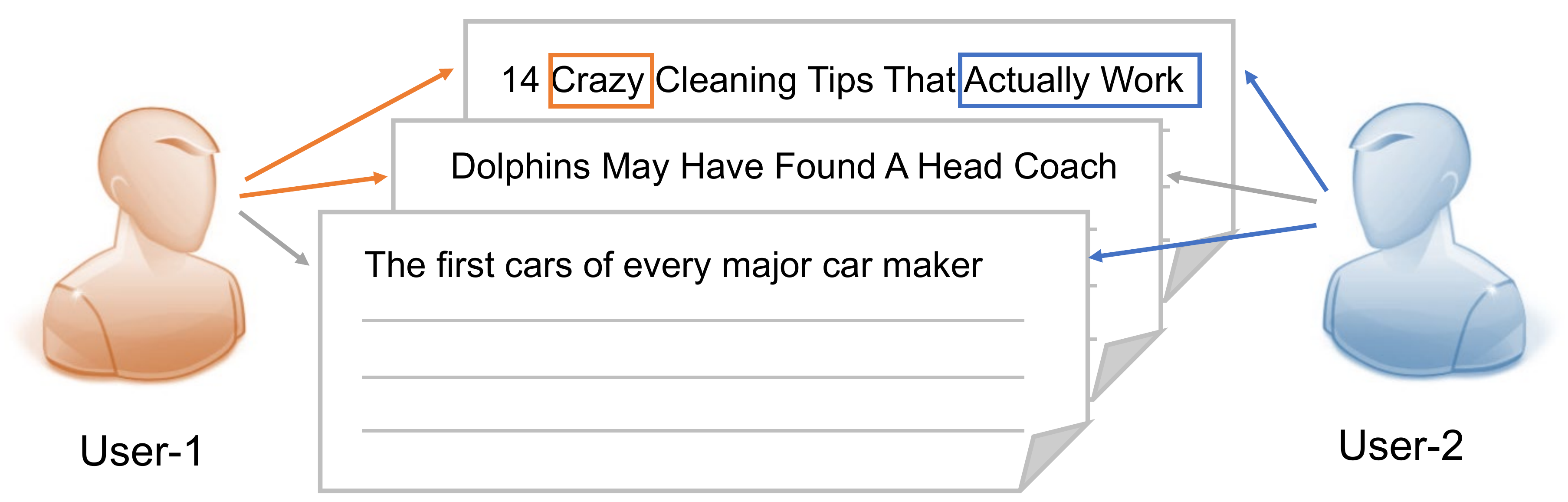}
  \caption{An illustrative example of two users and their clicked news articles. The colored arrows and boxes represent their interested news and words respectively.}
  \label{fig.example}
\end{figure}

There are two common observations in the news recommendation scenario.
First, not all news clicked by users can reflect the preferences of users.
For example, as illustrated in Figure~\ref{fig.example}, user-1 clicked all the three news, but he/she was only interested in the first and the second news.
In addition, the same news should also have different informativeness for modeling different users.
For example, if user-1 is very interested in sports news but user-2 rarely reads, the news ``Dolphins May Have Found A Head Coach'' is very informative for characterizing user-1, but less informative for user-2.
Second, different words in news titles usually have different informativeness for learning news representations.
For example, the word ``Crazy'' in the first news title is informative, while the word ``That'' is uninformative.
Moreover, the same words within a news title may also have different informativeness for revealing preferences of different users.
For example, user-1 may be attracted by the word ``Crazy'', and user-2 may pay more attention to the words ``Actually Work''.
Therefore, modeling the different informativeness of words and news for different users may be useful for learning better representations of users for accurate news recommendation.

Existing news recommendation methods are usually based on collaborative filtering (CF) techniques and news content\cite{liu2010personalized,lu2015content,okura2017embedding,lian2018towards}.
For example, Liu et al.~\cite{liu2010personalized} proposed a CF-based approach for news recommendation based on user interests.
They use a Bayesian model to extract the interest features of users based on the click distributions on news articles in different categories.
Okura et al.~\cite{okura2017embedding} proposed to first learn the distributed representations of news articles based on similarity and then use recurrent neural networks to learn user representations from browsing histories for click prediction.
Lian et al.~\cite{lian2018towards} proposed a deep fusion model (DMF) to learn representations of users and news using combinations of fully connected networks with different depth.
They also used attention mechanism to select important user features.
However, all these existing methods cannot model the different informativeness of news and their words for different users, which may be useful for improving the quality of personalized news recommendation.

In this paper, we propose a neural approach with personalized attention (NPA) for news recommendation.
The core of our approach is a news representation model and a user representation model.
In the news representation model we use a CNN network to learn the contextual representations of news titles, and in the user representation model we learn representations of users from their clicked news.
Since different words and news articles usually have different informativeness for learning representations of news and users, we propose to apply attention mechanism at both word- and news-level to select and highlight informative words and news. 
In addition, since the informativeness of the same words and news may be  different for different users, we propose a personalized attention network by using the embedding of user ID as the query vector of the word- and news-level attention networks to differentially attend to important words and news according to user preferences.
Extensive experiments on a real-world  dataset collected from MSN news validate the effectiveness of our approach on news recommendation.

\section{Related Work}\label{sec:RelatedWork}
\subsection{News Recommendation}
News recommendation is an important task in the data mining field, and have been widely explored over years.
Traditional news recommendation methods usually based on news relatedness~\cite{lv2011learning}, semantic similarities~\cite{capelle2012semantics} and human editors' demonstration~\cite{wang2017dynamic}.
However, the preferences of users cannot be effectively modeled.
Therefore, most news recommendation methods are based on CF techniques.
The earliest study on CF methods for news recommendation is the Grouplens project~\cite{konstan1997grouplens}, which applied CF methods to aggregate news from Usenet.
However, pure CF methods usually suffer from the sparsity and the cold-start
problems, which are especially significant in news recommendation scenarios~\cite{li2011scene}.
Thus, content-based techniques are usually complementary methods to CF~\cite{liu2010personalized,son2013location,bansal2015content,phelan2009using,okura2017embedding,lian2018towards,wang2018dkn,zheng2018drn}.
For example, Liu et al.~\cite{liu2010personalized} proposed to incorporate user interests for news recommendation.
They use a Bayesian model to predict the interests of users based on the distributions of their clicked news articles in different categories.
Okura et al.~\cite{okura2017embedding} proposed to learn news embeddings based on the similarities between news articles in the same and different categories. 
They use recurrent neural networks to learn user representations from the browsing histories through time to predict the score of news.
Lian et al.~\cite{lian2018towards} proposed a deep fusion model (DMF) to learn representations of users from various features extracted from their 
news reading, general web browsing and web searching histories.
They used an inception network with attention mechanism to select important user features and combine the features learned by networks with different depths.
Wang et al.~\cite{wang2018dkn} proposed to use the embeddings of the entities extracted from a knowledge graph as a separate channel of the CNN input.
However, these existing methods cannot simultaneously model the informativeness of words and news.
Different from all these methods, we propose to use personalized attention mechanism at both word- and new-level to dynamically recognize different informative words and news according to the preference of different users.
Experimental results validate the effectiveness of our approach.

\subsection{Neural Recommender Systems}
In recent years, deep learning techniques have been widely used in  recommender systems~\cite{wang2015collaborative}.
For example, Xue et al.~\cite{xue2017deep} proposed to use multi-layer neural networks to learn the latent factors of users and items in matrix factorization.
However, the content of users and items cannot be exploited, which is usually important for recommender systems.
Different from using neural networks within traditional matrix factorization frameworks, many methods apply neural networks to learn representations of users and items from raw features~\cite{huang2013learning,elkahky2015multi,cheng2016wide,he2017neural,guo2017deepfm,covington2016deep}.
For example, Huang et al.~\cite{huang2013learning} proposed a deep structured semantic model (DSSM) for click-through rate (CTR) prediction.
They first hashed the very sparse bag-of-words vectors into low-dimensional feature vectors based on character n-grams,
then used multi-layer neural networks to learn the representations of query and documents, and jointly predicted the click score of multiple documents.
Cheng et al.~\cite{cheng2016wide} proposed a Wide \& Deep approach to combine a wide channel using a linear transformer with a deep channel using multi-layer neural networks.
Guo et al.~\cite{guo2017deepfm} proposed a DeepFM approach which combines factorization machines with deep neural networks.
The two components share the same input features and the final predicted score is calculated from the combination of the output from both components.
However, these methods usually rely on hand-crafted features, and the dimension of feature vectors is usually huge.
In addition, they cannot effectively recognize the important contexts when learning news and user representations.
Different from the aforementioned methods, our approach can dynamically select important words and news for recommendation based on user preferences, which may be useful for learning more informative user representations for personalized news recommendation.

\section{Our Approach}\label{sec:Model}
\begin{figure*}[!t]
  \centering
    \includegraphics[width=0.90\linewidth]{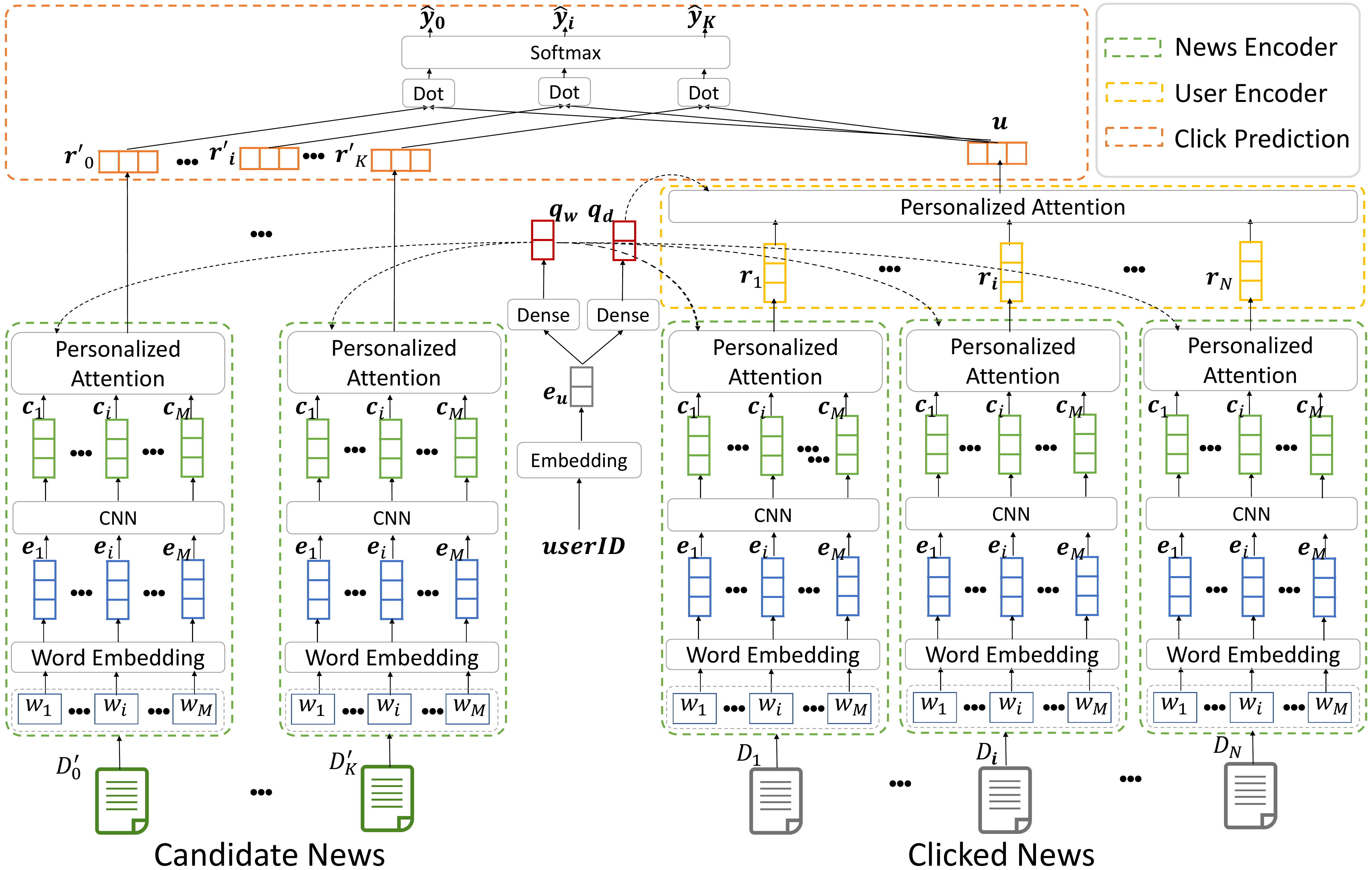}
  \caption{The framework of our NPA approach for news recommendation.}
  \label{fig.model}
\end{figure*}

In this section, we introduce our \textit{NPA} approach with personalized attention for news recommendation.
There are three major modules in our model.
The first one is a \textit{news encoder}, which aims to learn the representations of news.
The second one is a \textit{user encoder}, which aims to learn user representations based on the representations of his/her clicked news.
The third one is a \textit{click predictor}, which is used to predict the click score of a series of candidate news.
In the \textit{news encoder} and \textit{user encoder} module, we apply personalized attention networks at both word- and new-level to differentially select informative words and news according to user preferences.
The architecture of our approach is shown in Figure~\ref{fig.model}.
We will introduce the details of our approach in the following sections.

\subsection{News Encoder}
Since users' click decisions on news platforms are usually made based on the titles of news articles, in our approach the \textit{news encoder} module aims to learn news representations from news titles.
As shown in Figure~\ref{fig.model}, there are three sub-modules in the \textit{news encoder} module.

The first one is word embedding.
It is used to convert a sequence of words within news title into a sequence of low-dimensional dense vectors.
We denote the word sequence of the news $D_i$ as $D_i=[w_1,w_2,...,w_M]$, where $M$ is the number of words in the news title.
The word sequence $D_i$ is transformed into a sequence of vector $\textbf{E}^w=[\mathbf{e}_1,\mathbf{e}_2,...,\mathbf{e}_M]$ using a word embedding matrix $\mathbf{W}_e\in \mathcal{R}^{V\times D}$, where $V$ denotes the vocabulary size and $D$ denotes the dimension of word embedding vectors.

The second one is a convolutional neural network (CNN)~\cite{kim2014convolutional}.
CNN is an effective neural architecture for capturing local information~\cite{wu2019neuraldemo}.
Usually, local contexts within news are important for news recommendation.
For example, in the news title ``best Fiesta bowl moments'', the local combination of the words ``Fiesta'' and ``bowl'' is very important to infer the topic of this news.
Thus, we apply a CNN network to the word sequences to learn contextual representations of words within news titles by capturing their local contexts.
Denote the representation of the $i$-th word as $\mathbf{c}_i$, which is calculated as:
\begin{equation}
\mathbf{c}_i = \mathrm{ReLU}(\mathbf{F}_w\times \mathbf{e}_{(i-k):(i+k)}+\mathbf{b}_w),
\end{equation} 
where $\mathbf{e}_{(i-k):(i+k)}$ denotes the concatenation of the word embeddings from the position $(i-k)$ to $(i+k)$.
$\mathbf{F}_w \in \mathcal{R}^{N_f\times (2k+1)D}$ and $\mathbf{b}_w \in \mathcal{R}^{N_f}$ denote the parameters of the CNN filters, where $N_f$ is the number of CNN filters and $2k+1$ is their window size.
ReLU~\cite{glorot2011deep} is used as the non-linear function for activation.
The output of the CNN layer is the sequence of contextual word representation vectors, denoted as $[\mathbf{c}_1,\mathbf{c}_2,...,\mathbf{c}_M]$.

\begin{figure}[!t]
  \centering
    \includegraphics[width=0.5\linewidth]{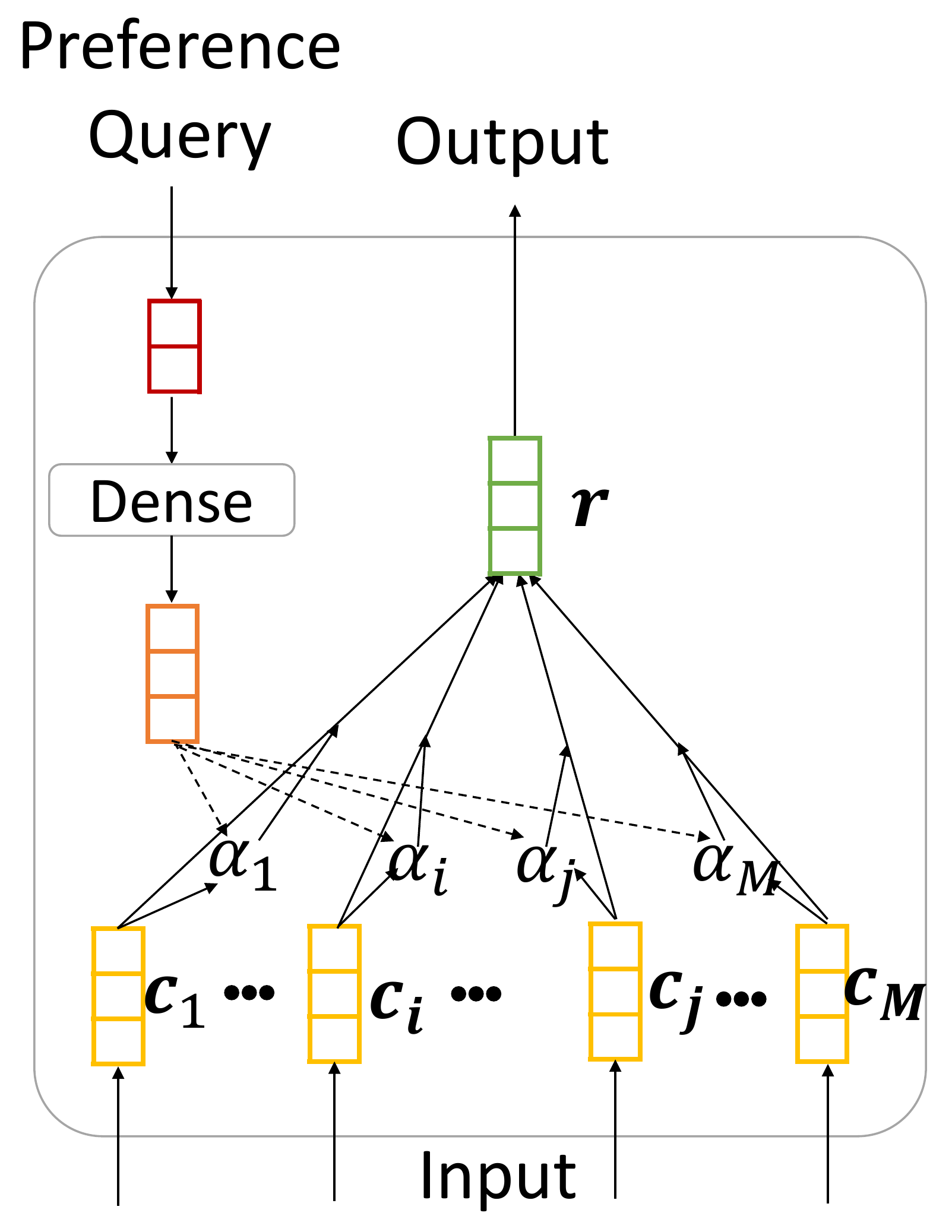}
 
   \caption{The architecture of the personalized attention module in our \textit{NPA} approach.}
  \label{fig.pa}

\end{figure}

The third one is a word-level personalized attention network.
Different words in a news title usually have different informativeness for characterizing the topics of news.
For example, in the news entitled with ``NBA games in this season'' the word ``NBA'' is very informative for learning news representations, since it conveys important clues about the news topic, while the word ``this'' is less informative for recommendation.
In addition, the same word may also have different informativeness for the recommendation of different users.
For example, in the news title ``Genesis G70 is the 2019 MotorTrend Car of the Year'', the words ``Genesis'' and ``MotorTrend'' are informative for the recommendation of users who are interested in cars, but may be less informative for users who are not interested.
Thus, recognizing important words for different users is useful for news recommendation.
However, in vanilla non-personalized attention networks~\cite{lian2018towards}, the attention weights are only calculated based on the input representation sequence via a fixed attention query vector, and the user preferences are not incorporated.
To model the informativeness of each word for the recommendation of different users, we propose to use a personalized attention network to recognize and highlight important words within news titles  according to user preferences. 
The architecture of our personalized attention module is shown in Figure~\ref{fig.pa}.
In order to obtain the representation of user preferences, we first embed the ID of users into a representation vector $\mathbf{e}_u \in \mathcal{R}^{D_e}$, where $D_e$ denotes the size of user embedding.
Then we use a dense layer to learn the word-level user preference query $\mathbf{q}_w$ as:
\begin{equation}
\mathbf{q}_w = \mathrm{ReLU}(\mathbf{V}_w \times \mathbf{e}_u+\mathbf{v}_w),
\end{equation}
where $\mathbf{V}_w \in \mathcal{R}^{D_e\times  D_q}$ and $\mathbf{v}_w\in \mathcal{R}^{ D_q}$ are parameters, $D_q$ is the preference query size.

In this module, the attention weight of each word is calculated based on the interactions between the preference query and word representations.
We denote the attention weight of the $i$-th word as $\alpha_i$, which is formulated as:
\begin{equation}
a_i=  \mathbf{c}_i^T\tanh(\mathbf{W}_p \times \mathbf{q}_w+\mathbf{b}_p),
\end{equation}
\begin{equation}
\alpha_i= \frac{\exp(a_i)}{\sum_{j=1}^M{\exp(a_j)}},
\end{equation}
where $\mathbf{W}_p \in \mathcal{R}^{D_q\times N_f}$ and $b_p \in \mathcal{R}^{N_f}$ are projection parameters.
The final contextual representation $\mathbf{r}_i$ of the $i$-th news title is the summation of the contextual representations of words weighted by their attention weights:
\begin{equation}
\mathbf{r}_i = \sum_{j=1}^M{\alpha_j  \mathbf{c}_j}.
\end{equation}
We apply the \textit{news encoder} to all users' clicked news and candidate news.
The representations of clicked news of a user and candidate news are respectively denoted as $[\mathbf{r}_1, \mathbf{r}_2, ..., \mathbf{r}_N]$ and $[\mathbf{r}'_0, \mathbf{r}'_1, ..., \mathbf{r}'_K]$, where $N$ is the number of clicked news and $K+1$ is the number of candidate news.
\subsection{User Encoder}
The \textit{user encoder} module in our approach aims to learn the representations of users from the representations of their clicked news, as shown in Figure~\ref{fig.model}.
In this module, a news-level personalized attention module is used to build informative user representations.
Usually the news clicked by the same user have different informativeness for learning user representations.
For example, the news ``10 tips for cooking'' is very informative for modeling user preferences, but the news ``It will be Rainy next week'' is less informative.
In addition, the same news also has different informativeness for modeling different users.
For example, the news ``100 Greatest Golf Courses'' is informative for characterizing users who are interested in golf, but is noisy for modeling users who are actually not interested in.
To model the different informativeness of the same news for different users, we also apply personalized attention mechanism to the representations of news clicked by the same user.
Similar with the word-level attention network, we first transform the user embedding vector into a news preference query $\mathbf{q}_d$, which is formulated as:
\begin{equation}
\mathbf{q}_d = \mathrm{ReLU}(\mathbf{V}_d \times \mathbf{e}_u+\mathbf{v}_d),
\end{equation}
where $\mathbf{V}_d \in \mathcal{R}^{D_e\times  D_d}$ and $\mathbf{v}_d\in \mathcal{R}^{D_d}$ are parameters, $D_d$ is the dimension of the news preference query.

We denote the attention weight of the $i$-th news as $\alpha'_i$, which is calculated by evaluating the importance of the interactions between the news preference query and news representations as follows:
\begin{equation}
a'_i=  \mathbf{r}_i^T\tanh(\mathbf{W}_d \times \mathbf{q}_d+b_d),
\end{equation}
\begin{equation}
\alpha'_i= \frac{\exp(a'_i)}{\sum_{j=1}^N{\exp(a'_j)}},
\end{equation}
where $\mathbf{W}_d \in \mathcal{R}^{D_d\times  N_f}$ and $b_d \in \mathcal{R}^{N_f}$ are projection parameters.
The final user representation $\mathbf{u}$ is the summation of the news contextual representations weighted by their attention weights:
\begin{equation}
\mathbf{u} = \sum_{j=1}^N{\alpha'_j  \mathbf{r}_j}.
\end{equation}

\subsection{Click Predictor}

The \textit{click predictor} module is used to predict the click score of a user on each candidate news.
A common observation in news recommendation is that most users usually only click a few news displayed in an impression.
Thus, the number of positive and negative news samples is highly imbalanced.
In many neural news recommendation methods~\cite{lian2018towards,wang2018dkn}, the model only predicts the click score for a single piece of news (the sigmoid activation function is usually used in these methods).
In these methods, positive and negative news samples are manually balanced by randomly sampling, and the rich information provided by negative samples is lost.
In addition, since the total number of news samples is usually huge, the computational cost of these methods is usually heavy during model training.
Thus, these methods are sub-optimal for simulating real-world news recommendation scenarios.
Motivated by~\cite{huang2013learning} and~\cite{zhai2016deepintent}, we propose to apply negative sampling techniques by jointly predicting the click score for $K+1$ news during model training to solve the two problems above.
The $K+1$ news consist of one positive sample of a user, and  $K$ randomly selected negative samples of a user.
The score $\hat{y}_i$ of the candidate news $D'_i$ is calculated by the inner product of the news and user representation vector first, and then normalized by the softmax function, which is formulated as:
\begin{equation}
\hat{y}'_i = \mathbf{r}_i^{'T}\mathbf{u},
\end{equation}
\begin{equation}
\hat{y}_i = \frac{\exp(\hat{y}'_i)}{\sum_{j=0}^K{\exp(\hat{y}'_j)}}
\end{equation}
For model training, we formulate the click prediction problem as a pseudo $K+1$ way classification task, i.e., the clicked news is the positive class and all the rest news are negative classes.
We apply maximum likelihood method to minimize the log-likelihood on the positive class:
\begin{equation}
 \mathcal{L}=-\sum_{y_j\in \mathcal{S}}\log(\hat{y}_j),
 \label{eq}
\end{equation}
where $y_j$ is the gold label, $\mathcal{S}$ is the set of the positive training samples.
By optimizing the loss function $\mathcal{L}$ via gradient descend, all parameters can be tuned in our model.
Compared with existing news recommendation methods, our approach can effectively exploit the useful information in negative samples, and further reduce the computational cost for model training (nearly divided by $K$).
Thus, our model can be trained more easily on a large collection of news click logs.

\section{Experiments}\label{sec:Experiments}

\subsection{Datasets and Experimental Settings}

Our experiments were conducted on a real-world dataset, which was constructed by randomly sampling user logs from MSN News\footnote{https://www.msn.com/en-us/news} in one month, i.e., from December 13rd, 2018 to January 12nd, 2019.
The detailed statistics of the dataset is shown in Table~\ref{dataset}\footnote{All words are lower-cased.}.
We use the logs in the last week as the test set, and the rest logs are used for training.
In addition, we randomly sampled 10\% of samples for validation.

In our experiments, the dimension $D$ of word embedding was set to 300.
we used the pre-trained Glove embedding\footnote{http://nlp.stanford.edu/data/glove.840B.300d.zip}~\cite{pennington2014glove}, which is trained on a corpus with 840 billion tokens, to initialize the embedding matrix. 
The number of CNN filters $N_f$ was set to 400, and the window size was 3.
The dimension of user embedding $D_e$ was set to 50.
The sizes of word and news preferences queries ($D_q$ and $D_d$) were both set to 200.
The negative sampling ratio $K$ was set to 4.
We randomly sampled at most 50 browsed news articles to learn user representations.
We applied dropout strategy~\cite{srivastava2014dropout} to each layer in our approach to mitigate overfitting.
The dropout rate was set to 0.2.
Adam~\cite{kingma2014adam} was used as the optimization algorithm for gradient descend.
The batch size was set to 100.
Due to the limitation of GPU memory, the maximum number of clicked news for learning user representations was set to 50 in neural network based methods, and the maximum length of news title was set to 30.
These hyperparameters were selected according to the validation set.
The metrics in our experiments include the average AUC, MRR, nDCG@5 and nDCG@10 scores over all impressions.
We independently repeated each experiment for 10 times and reported the average performance.

\begin{table}[h]
\caption{Statistics of our dataset. *Denote the ratio of the negative sample number to positive sample number.}
\label{dataset}
	\resizebox{0.48\textwidth}{!}{
\begin{tabular}{|c|r|c|r|}
\hline
\textbf{\# users}       & 10,000     & \textbf{avg. \# words per title} & 11.29      \\ \hline
\textbf{\# news}        & 42,255     & \textbf{\# positive samples}     & 489,644 \\ \hline
\textbf{\# impressions} & 445,230    & \textbf{\# negative samples}     & 6,651,940  \\ \hline
\textbf{\# samples}     & 7,141,584 & \textbf{NP ratio*}                & 13.59      \\ \hline
\end{tabular}
}
\end{table}

\subsection{Performance Evaluation}
First, we will evaluate the performance of our approach by comparing it with several baseline methods.
The methods to be compared include:
\begin{itemize}
    \item \textit{LibFM}~\cite{rendle2012factorization}, which is a state-of-the-art feature-based matrix factorization and it is a widely used method for recommendation. In our experiments, we extract the TF-IDF features from users' clicked news and candidate news, and concatenate both types of features as the input for LibFM.
    \item \textit{CNN}~\cite{kim2014convolutional}, applying CNN to the word sequences in news titles and use max pooling technique to keep the most salient features, which is widely used in content-based recommendation~\cite{zheng2017joint,chen2018neural}.
    \item DSSM~\cite{huang2013learning}, a deep structured semantic model with word hashing via character trigram and multiple dense layers.
    In our experiments, we concatenate all user’s clicked news as a long document as the query, and the candidate news are documents.
    The negative sampling ratio was set to 4.
    \item \textit{Wide \& Deep}~\cite{cheng2016wide}, using the combination of a wide channel using a linear transformer and a deep channel with multiple dense layers. 
    The features we used are the same with \textit{LibFM} for both channels.
    \item \textit{DeepFM}~\cite{guo2017deepfm}, which is also a widely used neural recommendation method, using a combination with factorization machines and deep neural networks.
    We use the same TF-IDF features to feed for both components.
    \item \textit{DFM}~\cite{lian2018towards}, a deep fusion model by using combinations of dense layers with different depths.
    We use both TF-IDF features and word embeddings as the input for DFM.
    \item \textit{DKN}~\cite{wang2018dkn}, a deep news recommendation method with Kim CNN and news-level attention network.
    They also incorporated entity embeddings derived from knowledge graphs.
    \item \textit{NPA}, our neural news recommendation approach with personalized attention.
\end{itemize}

\begin{table}[h]
	\centering
	
	\caption{The performance of different methods on news recommendation. *The improvement over all baseline methods is significant at the level $p<0.001$.}\label{table.result}
	\resizebox{0.48\textwidth}{!}{
\begin{tabular}{|c|c|c|c|c|}
\hline
    \textbf{Methods}         & \textbf{AUC}             & \textbf{MRR}             & \textbf{nDCG@5}          & \textbf{nDCG@10}         \\ \hline
LibFM        & 0.5660          & 0.2924          & 0.3015          & 0.3932          \\
CNN          & 0.5689          & 0.2956          & 0.3043          & 0.3955          \\
DSSM         & 0.6009          & 0.3099          & 0.3261          & 0.4185          \\ 
Wide \& Deep & 0.5735          & 0.2989          & 0.3094          & 0.3996          \\
DeepFM       & 0.5774          & 0.3031          & 0.3122          & 0.4019          \\ 
DFM          & 0.5860          & 0.3034          & 0.3175          & 0.4067          \\
DKN          & 0.5869          & 0.3044          & 0.3184          & 0.4071          \\ \hline
NPA         & \textbf{0.6243}*  & \textbf{0.3321}*  & \textbf{0.3535}*  & \textbf{0.4380}*  \\ \hline
\end{tabular}
}
\end{table}

The experimental results on news recommendation are summarized in Table~\ref{table.result}.
According to Table~\ref{table.result}, We have several observations.

First, the methods based on neural networks (e.g., \textit{CNN}, \textit{DSSM} and \textit{NPA}) outperform traditional matrix factorization methods such as \textit{LibFM}.
This is probably because neural networks can learn more sophisticated features than \textit{LibFM}, which is beneficial for learning more informative latent factors of users and news.

Second, the methods using negative sampling (\textit{DSSM} and \textit{NPA}) outperform the methods without negative sampling (e.g., \textit{CNN}, \textit{DFM} and \textit{DKN}).
This is probably because the methods without negative sampling are usually trained on a balanced dataset with the same number of positive and negative samples, and cannot effectively exploit the rich information conveyed by negative samples.
\textit{DSSM} and our \textit{NPA} approach can utilize the information from three more times of negative samples than other baseline methods, which is more suitable for real-world news recommendation scenarios.

Third, the deep learning methods using attention mechanism (\textit{DFM}, \textit{DKN} and \textit{NPA}) outperform most of the methods without attention mechanism (\textit{CNN}, \textit{Wide \& Deep} and \textit{DeepFM}).
This is probably because different news and their contexts usually have different informativeness for recommendation, and selecting the important features of news and users is useful for achieving better recommendation performance.

Fourth, our approach can consistently outperform all compared baseline methods.
Although \textit{DSSM} also use negative sampling techniques to incorporate more negative samples, it cannot effectively utilize the contextual information and word orders in news titles.
Thus, our approach can outperform \textit{DSSM}.
In addition, although \textit{DFM} uses attention mechanism to select important user features, it also cannot effectively model the contexts within news titles, and cannot select important words in the candidate news titles.
Besides, although \textit{DKN} uses a news-level attention network to select the news clicked by users, it cannot model the informativeness of different words.
Different from all these methods, our approach can dynamically recognize important words and news according to user preferences.
Thus, our approach can outperform these methods.

\begin{table}[!t]
	\caption{The time and memory complexities of different methods. $N_s$ is the number of training samples. $D_f$ and $d_f$ are the dimension of the feature vector and latent factors respectively. *The computational cost per sample.}\label{table.cost}
	\resizebox{0.48\textwidth}{!}{
\begin{tabular}{|c|c|c|c|c|}
\hline
\multirow{2}{*}{\textbf{Methods}} & \multicolumn{2}{c|}{\textbf{Training}}       & \multicolumn{2}{c|}{\textbf{Test*}} \\ \cline{2-5} 
                         & Time               & Memory         & Time       & Memory       \\ \hline
LibFM                    & $O(N_sD_f)$       & $O(N_sD_f)$   & $O(d_f)$   & $O(d_f)$     \\ 
CNN                      & $O(N_sND_e)$     & $O(N_sND_e)$ & $O(d_f)$   & $O(d_f)$     \\ 
DSSM                     & $O(N_sD_f/K)$     & $O(N_sD_f)$   & $O(d_f)$   & $O(d_f)$     \\ 
Wide \& Deep             & $O(N_sD_f)$       & $O(N_sD_f)$   & $O(d_f)$   & $O(d_f)$     \\
DeepFM                   & $O(N_sD_f)$       & $O(N_sD_f)$   & $O(d_f)$   & $O(d_f)$     \\ 
DFM                      & $O(N_s(ND+D_f))$ & $O(N_s(ND+D_f))$   & $O(d_f+D_f)$   & $O(d_f+D_f)$     \\
DKN                      & $O(N_sND)$       & $O(N_sND)$   & $O(d_fN)$ & $O(ND+d_f)$ \\ \hline
NPA                      & $O(N_sND/K)$     & $O(N_sND)$   & $O(d_f+D+D_e)$ & $O(d_f+D_e)$ \\ \hline
\end{tabular}
}
\end{table}
Next, we will compare the computational cost of our approach and the baseline methods.
To summarize, the comparisons are shown in Table~\ref{table.cost}\footnote{We omit the length of news titles since it is usually set to a constant. In addition, we also omit the dimensions of hidden layers because they are usually close to the word embedding dimension $D$.}.
We assume that during the online test phase, the model can  directly use the intermediate results produced by hidden layers.
From Table~\ref{table.cost}, we have several observations.

First, comparing our \textit{NPA} approach with feature-based methods, the computational cost on time and memory during training is lower if $N$ is not large, since the dimension $D_f$ of the feature vector is usually huge due to the dependency on bag-of-words features.\footnote{In practice, the training time of \textit{LibFM} is about 20 minutes, while \textit{DSSM}, \textit{Wide \& Deep}, \textit{DeepFM} all take more than one hour on a GTX1080ti GPU. Our \textit{NPA} only takes about 15 minutes.}
In addition, the computational cost in the test phase is only a little more expensive than these methods since $D_e$ is not large.

Second, comparing our \textit{NPA} approach with \textit{CNN}, \textit{DFM} and \textit{DKN}, the training cost is actually divided by $K$ with the help of negative sampling.
In addition, the computational cost of \textit{NPA} in the test phase much smaller than \textit{DKN} and \textit{DFM}, since \textit{DKN} needs to use the representations of the candidate news as the query of the news-level attention network and the score needs to be predicted by encoding all news clicked by a user, which is very computationally expensive. \textit{DFM} needs to take the sparse feature vector as input, which is also  computationally expensive.
Different from these baseline methods, our approach can be trained at a low computational cost, and can be applied to online services to handle massive users at a satisfactory computational cost.

\begin{figure}[!t]
  \centering
    \includegraphics[width=0.7\linewidth]{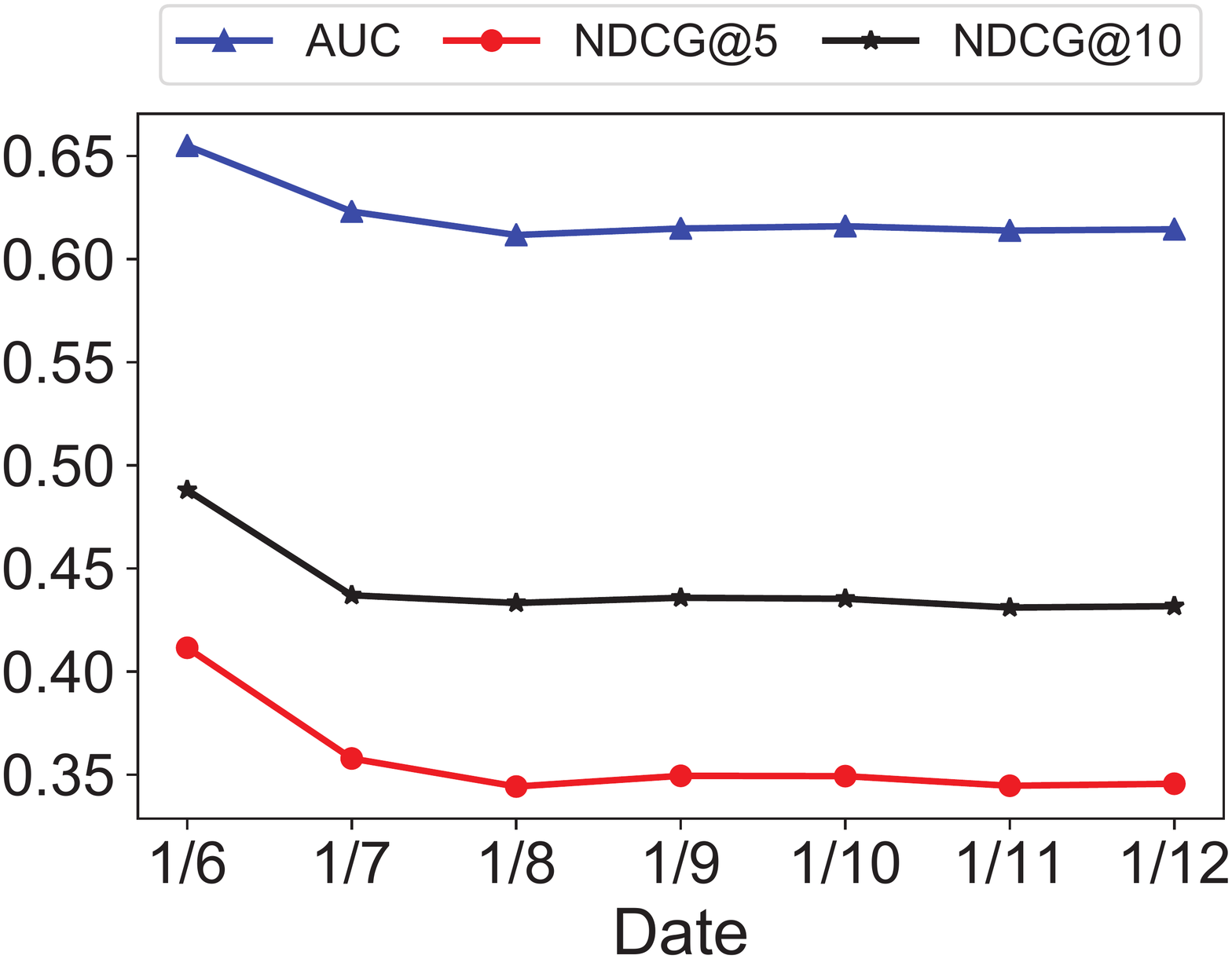}
  \caption{The performance of our \textit{NPA} approach in different days of a week (1/6/2019-1/12/2019).}
  \label{fig.time}
\end{figure}

Finally, we want to evaluate the performance of our approach in each day to explore the influence of user click behaviors over time.
The performance of our approach in each day of the week for test (1/6/2019-1/12/2019) is shown in Figure~\ref{fig.time}.
According to the results,
the performance of our approach is  best on the first day in the test week (1/6/2019).
This is intuitive because the relevance of user preferences is usually strong between neighbor days.
In addition, as time went by, the performance of our approach begins to decline.
This is probably because news are usually highly time-sensitive and most articles in common news services will be no longer recommended for users within several days (Usually two days for MSN news).
Thus, more news will not appear in the training set over time, which leads to the performance decline.
Fortunately, we also find the performance of our approach tends to be stable after three days.
It shows that our model does not simply memorize the news appear in the training set and can make personalized recommendations based on user preferences and news topics.
Thus, our model is robust to the news update over time.
\begin{figure}[!t]
  \centering
    \includegraphics[width=0.7\linewidth]{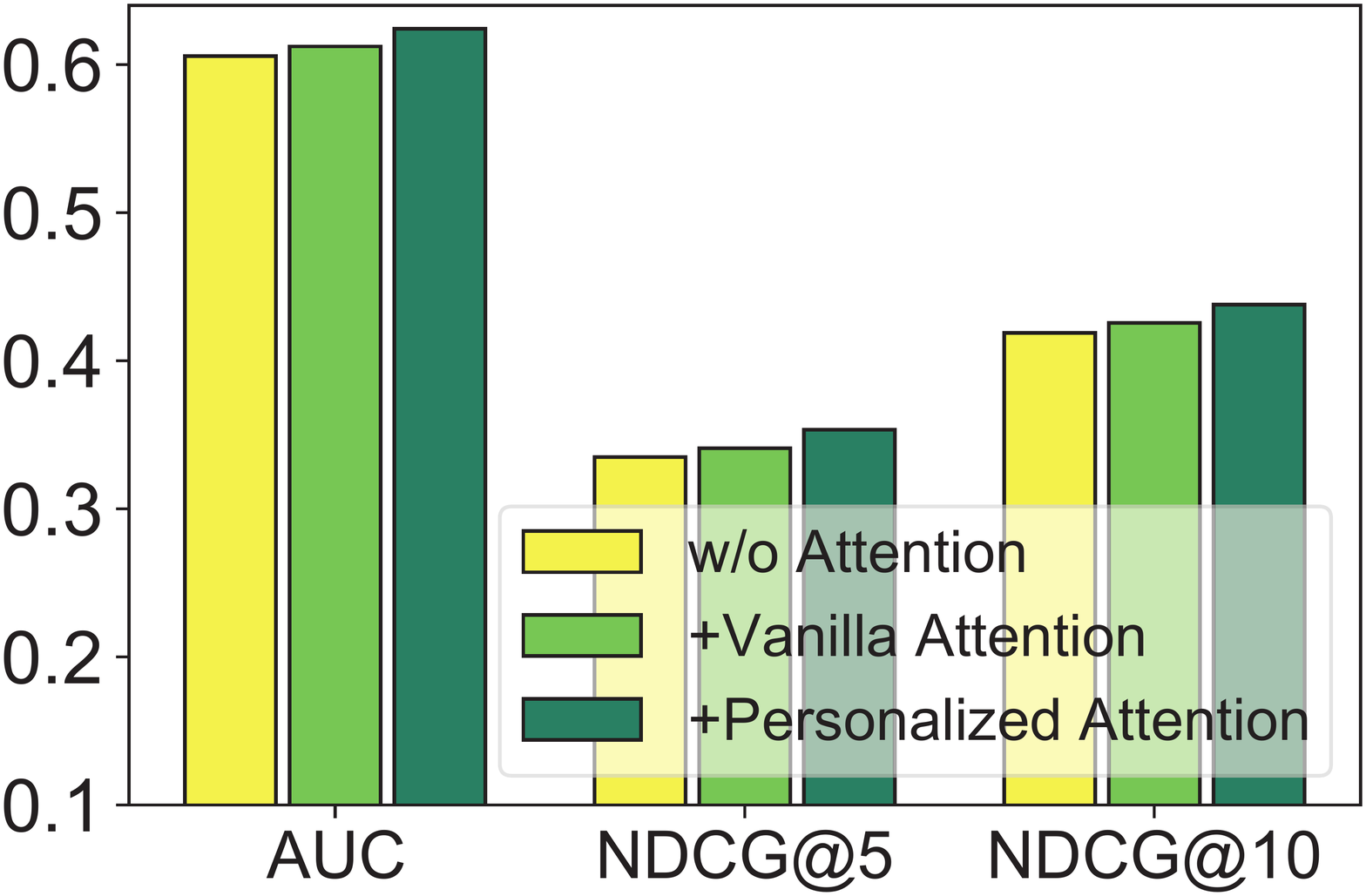}
  \caption{The effectiveness of our personalized attention mechanism compared with vanilla attention.}
  \label{fig.att1} 
\end{figure}
\begin{figure}[!t]
  \centering
    \includegraphics[width=0.65\linewidth]{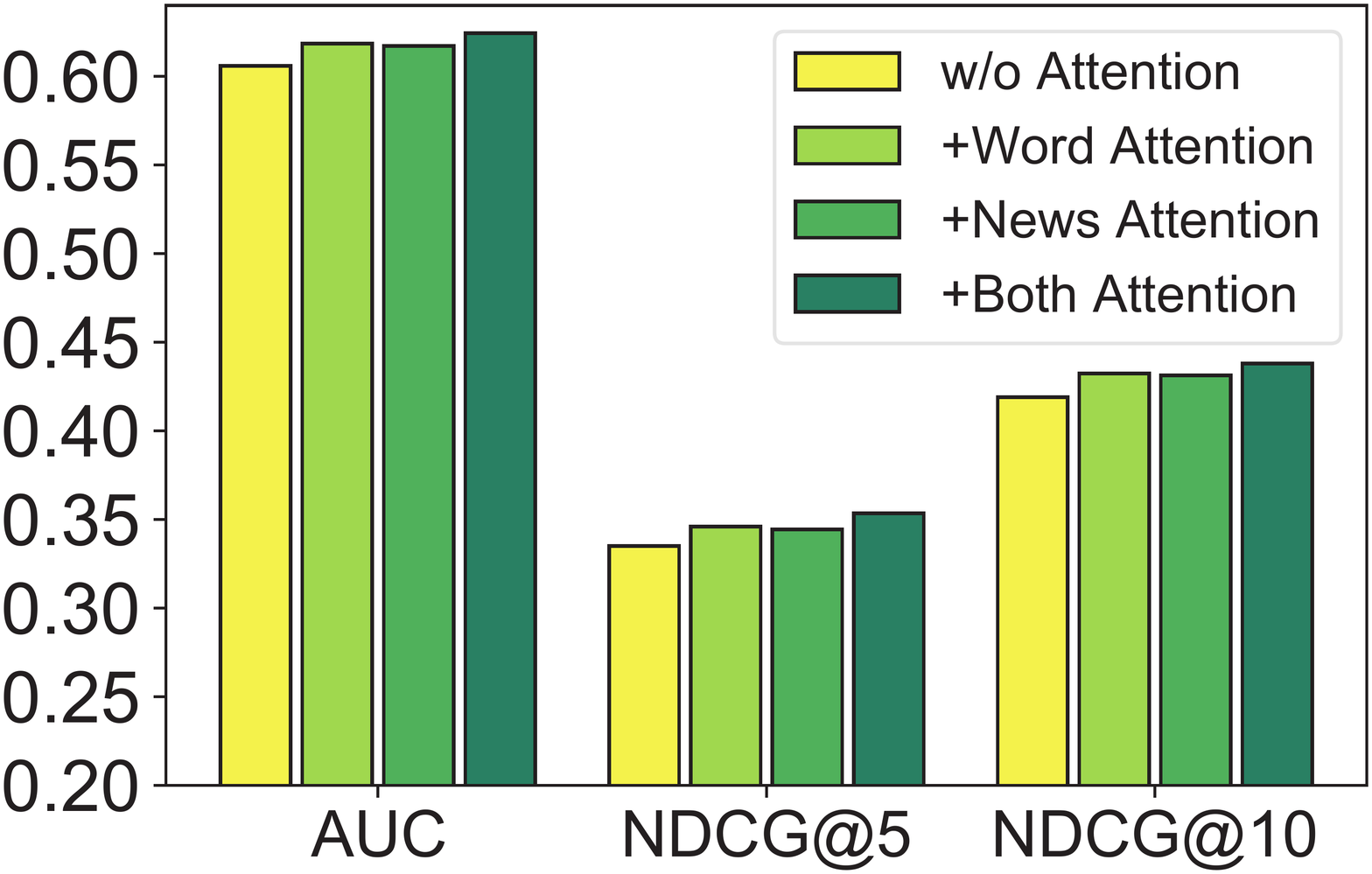}

  \caption{The effectiveness of the word-level and news-level personalized attention network.}
  \label{fig.att2} 
\end{figure}

\subsection{Effectiveness of Personalized Attention}

In this section, we conducted several experiments to explore the effectiveness of the personalized attention mechanism in our \textit{NPA} approach.
First, we want to validate the advantage of  personalized attention on vanilla non-personalized attention for news recommendation.
The performance of \textit{NPA} and its variant using vanilla attention and without attention is shown in Figure~\ref{fig.att1}.
According to Figure~\ref{fig.att1}, we have several observations.
First, the models with attention mechanism consistently outperform the model without attention.
This is probably because different words and news usually have different informativeness for news recommendation.
Therefore, using attention mechanism to recognize and highlight important words and news is useful for learning more informative news and user representations.
In addition, our model with personalized attention outperforms its variant with vanilla attention.
This is probably because the same words and news should have different informativeness for the recommendation of different users.
However, vanilla attention networks usually use a fixed query vector, and cannot adjust to different user preferences.
Different from vanilla attention, our personalized attention approach can dynamically attend to important words and news according to user characteristics, which can benefit news and user representation learning.

Then, we want to validate the effectiveness of the personalized attention at word-level and news-level.
The performance of \textit{NPA} and its variant with different combinations of personalized attention is shown in Figure~\ref{fig.att2}.
According to Figure~\ref{fig.att2}, we have several observations.
First, the word-level personalized attention can effectively improve the performance of our approach.
This is probably because words are basic units to convey the meanings of news titles and selecting the important words according to user preferences is useful for learning more informative news representations for personalized recommendation.
Second, the news-level personalized attention can also improve the performance of our approach.
This is probably because news clicked by users usually have different informativeness for learning user representations, and recognizing the important news is useful for learning high quality user representations.
Third, combining both word- and news-level personalized attention can further improve the performance of our approach.
These results validate the effectiveness of the personalized attention mechanism in our approach.

\begin{figure}[!t]
  \centering
    \includegraphics[width=0.65\linewidth]{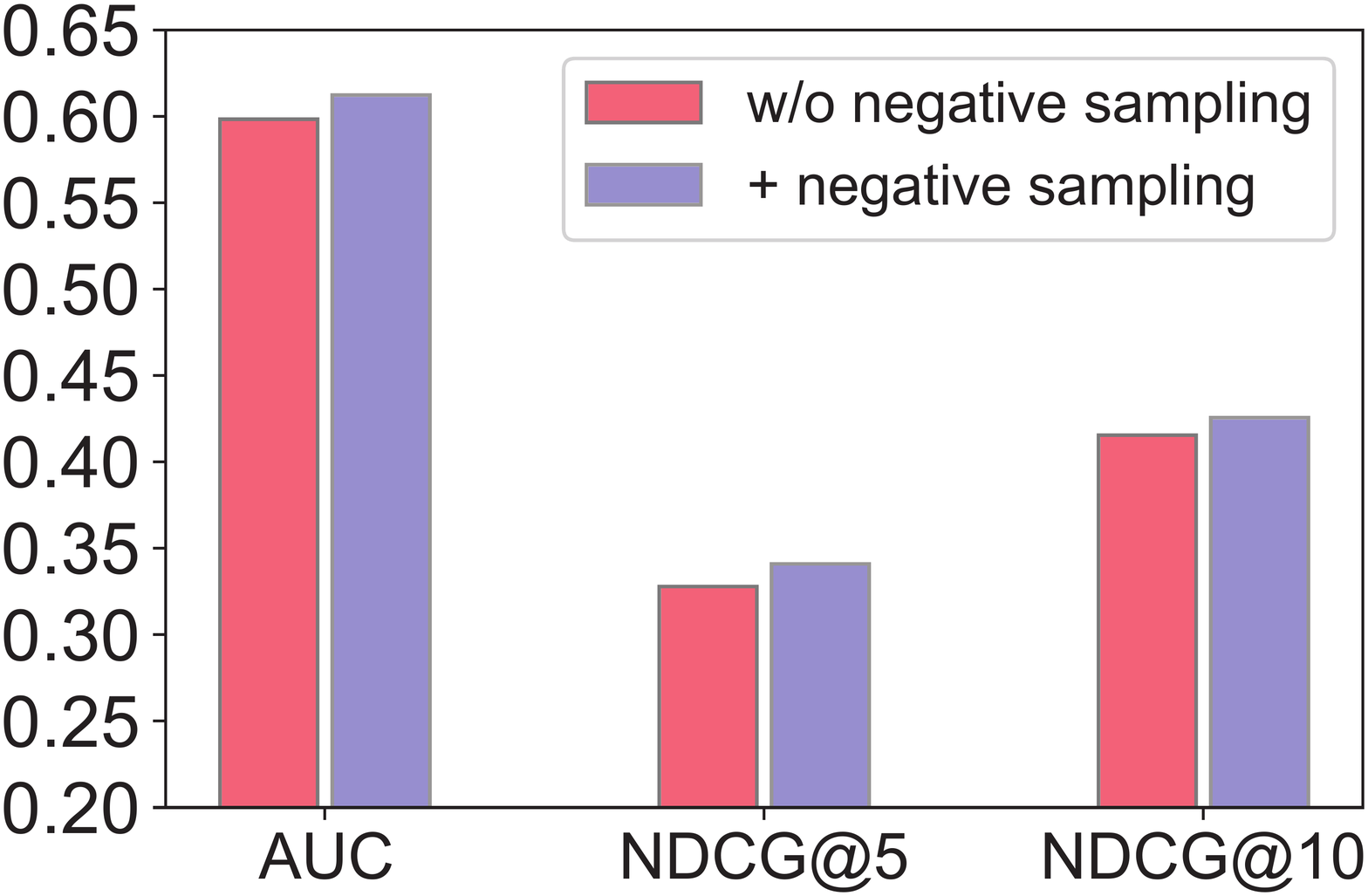}
  \caption{The influence of negative sampling on the performance of our approach.}
  \label{fig.negative}
\end{figure}
\begin{figure}[!t]
  \centering
    \includegraphics[width=0.7\linewidth]{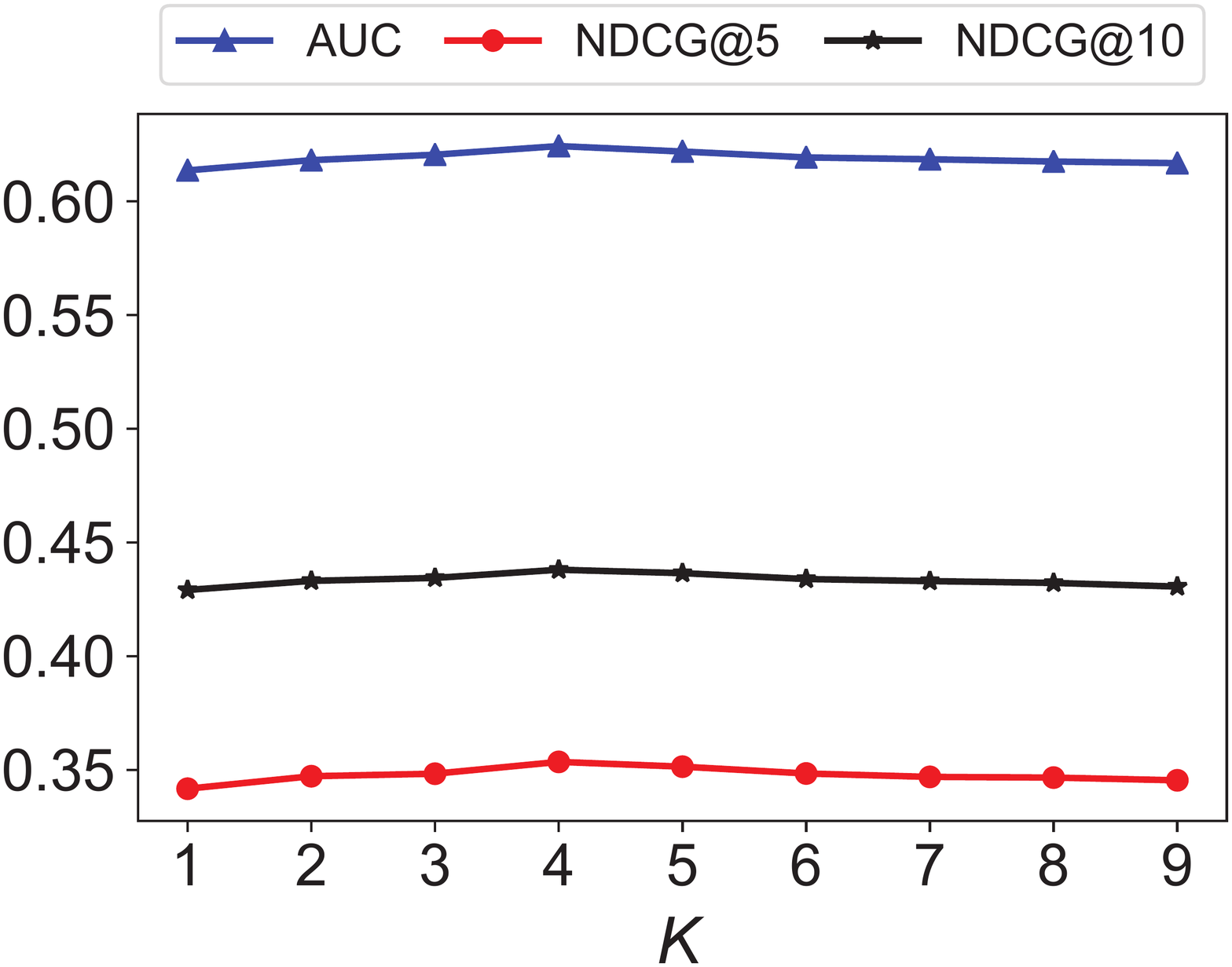}
  \caption{Influence of the negative sampling ratio $K$ on the performance of our approach.}
  \label{lambda}
\end{figure}

\begin{figure*}[t]
	\centering
	\subfigure[Word-level attention weights.]
	{
		\label{fig.word} 
		\includegraphics[width=0.85\textwidth]{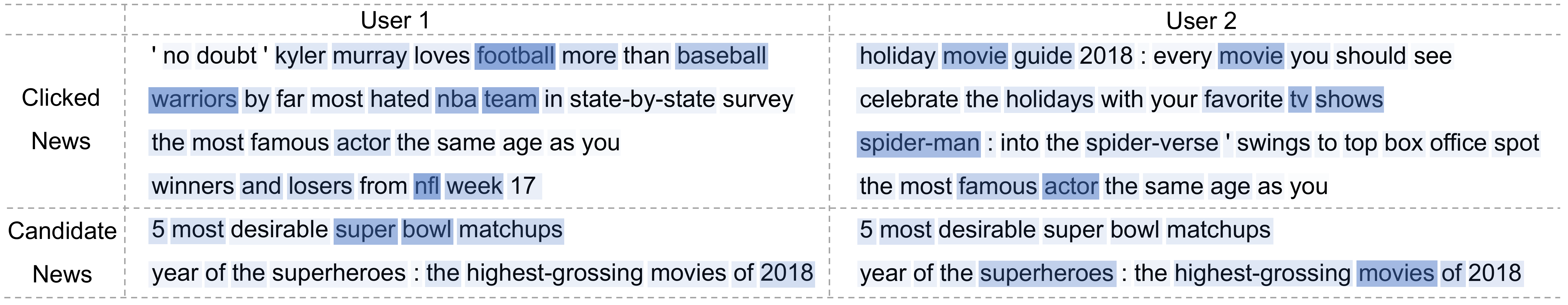}
	}
	\subfigure[News-level attention weights.]
	{
		\label{fig.news} 
		\includegraphics[width=0.85\textwidth]{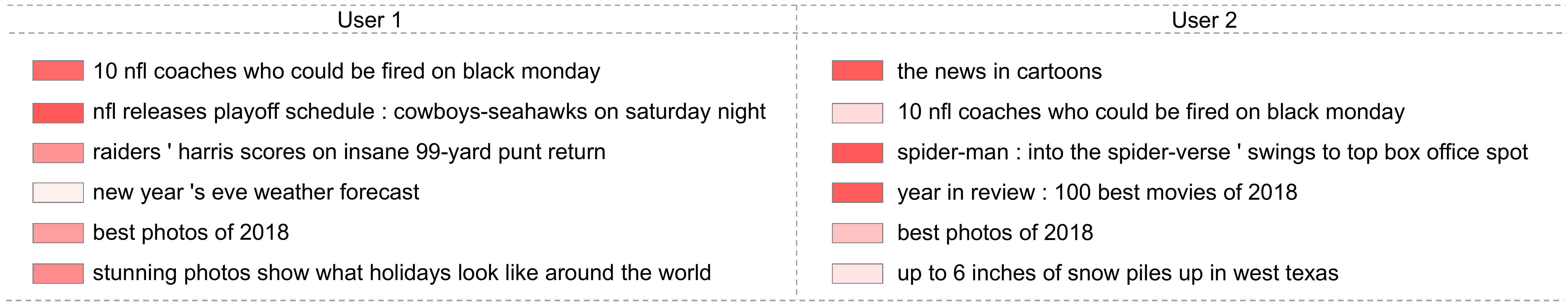}
	}
	\caption{Visualization of the attention weights from the word- and news-level personalized attention network.
  The users and news are randomly sampled from the dataset. Darker colors indicate higher attention weights.}\label{fig.case}
\end{figure*}

\subsection{Influence of Negative Sampling}
In this section, we will explore the influence of the negative sampling technique on the performance of our  approach.
To validate the effectiveness of negative sampling, we compare the performance of our approach with its variant without negative sampling.
Following~\cite{wang2018dkn,lian2018towards}, we choose to train this variant  on a balanced training set by predicting the click scores of news articles one by one (the final activation function is changed to sigmoid).
The experimental results are shown in Figure~\ref{fig.negative}.
According to Figure~\ref{fig.negative}, the performance of our approach can be effectively improved via negative sampling.
Since negative samples are dominant in the training set, they usually contain rich information for recommendation.
However, the information provided by negative samples cannot be effectively utilized due to the limitation of balanced sampling.
Therefore, the performance is usually sub-optimal.
Different from this variant, our \textit{NPA} approach can incorporate richer information in negative samples, which may be very useful for achieving better performance on news recommendation.

\subsection{Hyperparameter Analysis}

In this section, we will explore the influence of an important hyperparameter in our approach, i.e., the negative sampling ratio $K$, which aims to control the number of negative samples to combine with a positive sample.
The experimental results on $K$ are shown in Figure~\ref{lambda}.
According to Figure~\ref{lambda}, we find the performance of our approach first consistently improves when $K$ increases.
This is probably because when $K$ is too small, the number of negative samples incorporated for training is also small, and  the useful information provided by negative samples cannot be fully exploited, which will lead to sub-optimal performance.
However, when $K$ is too large, the performance will start to decline.
This is probably because when too many negative samples are incorporated, they may become dominant and it is difficult for the model to correctly recognize the positive samples, which will also lead to sub-optimal performance.
Thus, a moderate setting of $K$ may be more appropriate (e.g., $K=4$).

\subsection{Case Study}
In this section, we will conduct several case studies to visually explore the effectiveness of the personalized attention mechanism in our approach.
First, we want to explore the effectiveness of the word-level personalized attention.
The visualization results of the clicked and candidate news from two sample users are shown in Figure~\ref{fig.word}.
From Figure~\ref{fig.word}, we find the attention network can effectively recognize important words within news titles.
For example, the word ``nba'' in the second news of user 1 is assigned high attention weight since it is very informative for modeling user preferences, while the word ``survey'' in the same title gains low attention weight since it is not very informative.
In addition, our approach can calculate different attention weights for the words in the same news titles to adjust to the preferences of different users.
For example, according to the clicked news, user 1 may be interested in sports news and user 2 may be interested in movie related news.
The words ``super bowl'' are highlighted for user 1 and the words ``superheroes'' and ``movies'' are highlighted for user 2.
These results show that our approach can learn personalized news representations by incorporating personalized attention.

Then, we want to explore the effectiveness of the news-level attention network.
The visualization results of the clicked news are shown in Figure~\ref{fig.news}.
From Figure~\ref{fig.news}, we find our approach can also effectively recognize important news of a user.
For example, the news ``nfl releases playoff schedule : cowboys-seahawks on saturday night'' gains high attention weight because it is very informative for modeling the preferences of user 1, since he/she is very likely to be interested in sports according to the clicked news.
The news ``new year's eve weather forecast'' is assigned low attention weight, since it is uninformative for modeling user preferences.
In addition, our approach can model the different informativeness of news for learning representations of different users.
For example, the same sports news ``10 nfl coaches who could be fired on black monday'' is assigned high attention weight for user 1, but relatively low for user 2.
According to the clicked news of both users, user 1 is more likely to be interested in sports than user 2 and this news may be noisy for user 2.
These results show that our model can evaluate the different importance of the same news for different users according to their preferences.

\section{Conclusion}\label{sec:Conclusion}

In this paper, we propose a neural news recommendation approach with personalized attention (NPA).
In our \textit{NPA} approach, we use a  news representation model to learn news representations from titles using CNN, and use a user representation model to learn representations of users from their clicked news.
Since different words and news articles usually have different informativeness for representing news and users, we proposed to apply attention mechanism at both word- and news to help our model to attend to important words and news articles.
In addition, since the same words and news  usually have different importance for different users, we propose a personalized attention network which exploits the embeddings of user ID as the queries of  the word- and news-level attention networks.
The experiments on the real-world dataset collected from MSN news validate the effectiveness of our approach.

\section*{Acknowledgments}
The authors would like to thank Microsoft News for providing technical support and data in the experiments, and Jiun-Hung Chen (Microsoft News) and Ying Qiao (Microsoft News) for their support and discussions.
This work was supported by the National Key Research and Development Program of China under Grant number 2018YFC1604002, the National Natural Science Foundation of China under Grant numbers U1836204, U1705261, U1636113, U1536201, and U1536207, and the Tsinghua University Initiative Scientific Research Program under Grant number 20181080368.

\bibliographystyle{ACM-Reference-Format}
\bibliography{weibo}
\end{document}